\documentclass[9pt,twocolumn,twoside]{opticajnl}
\journal{opticajournal} 

\setboolean{shortarticle}{true}

\usepackage{lineno}
\usepackage{amssymb}

\title{Higher-order dark solitons and oscillatory dynamics \\ in microcavity polariton condensates}
\author[1,2]{Jinming Sun}
\author[1,*]{Manna Chen}
\author[3,4,5]{Stefan Schumacher}
\author[2]{Wei Hu}
\author[3]{Xuekai Ma}

\affil[1]{School of Electrical Engineering and Intelligentization, Dongguan University of Technology, Dongguan 523808, China}
\affil[2]{Guangdong Provincial Key Laboratory of Nanophotonic Functional Materials and Devices, South China Normal University, Guangzhou 510631, China}
\affil[3]{Department of Physics and Center for Optoelectronics and Photonics Paderborn (CeOPP),  Paderborn University, 33098 Paderborn, Germany}
\affil[4]{Institute for Photonic Quantum Systems (PhoQS), Paderborn University, 33098 Paderborn, Germany}
\affil[5]{Wyant College of Optical Sciences, University of Arizona, Tucson, Arizona 85721, USA}

\affil[*]{Corresponding author: chenmn@dgut.edu.cn}

\begin{abstract}
Dark solitons carrying quantized phase information arouse great interest in different nonlinear systems. A dark soliton in 1D can be stabilized in microcavity polariton condensates as a confinement is imposed on it to prevent its decay. Such a confinement can be realized by optical manners, i.e., by using optically induced potential traps. Under nonresonant excitation with spatially periodically modulated optical beams, we numerically demonstrate that besides fundamental dark solitons, higher-order dark solitons with multiple density minima and $\pi$-phase jumps can also stably survive in the potential (pump) valleys. Simultaneously exciting several orders of dark soliton states by properly choosing the lattice constant of the optical pump gives rise to dark oscillators. In some cases, the stably trapped dark solitons in adjacent pump valleys squeeze the condensate density between them and generate another type of density dips in the pump peak area. Surprisingly, such a density dip supports another stable dark soliton with a larger size which is essentially composed of two counter-propagating gray solitons.
\end{abstract}

\setboolean{displaycopyright}{false} 

\begin{document}

\maketitle

A dark soliton is a nonlinear phenomenon, characterized by a $\pi$-phase jump and accompanied by an amplitude dip on a continuous background, and has been intensively studied in many physical systems, such as in nonlinear optics~\cite{kivshar2003optical,trillo2013spatial}, atomic condensates~\cite{PhysRevLett.83.5198,denschlag2000generating,becker2008oscillations}, and light-matter (photon-exciton) hybrid exciton-polariton systems~\cite{amo2011polariton,PhysRevLett.107.245301,PhysRevB.83.193305,PhysRevB.86.020509,PhysRevB.89.235310,PhysRevLett.112.216401,PhysRevLett.112.140405,PhysRevLett.113.103901,dominici2015real,PhysRevLett.117.217401,PhysRevLett.118.157401,xue2018creation,PhysRevX.10.041028,PhysRevB.105.245302,PhysRevResearch.5.043062}. In nonlinear optics, dark optical solitons can occur in media with defocusing nonlinearites. In atomic condensates, dark solitons are allowed when interactions are repulsive. As hybrid quasi-particles, polariton condensates also present strong repulsive interactions that in principle permit the dark solitons to survive. Different from atomic condensates, polariton condensates are driven-dissipative systems and they can be created under nonresonant excitation with laser beams. The nonequilibrium nature of polariton condensates, however, destabilizes imprinted dark solitons with homogeneous background excitation, so that they quickly disappear or decay into vortex-antivortex pairs~\cite{PhysRevB.89.235310,PhysRevLett.112.216401}.

A possible way to stabilize dark solitons in polariton condensates is to spatially modulate the optical pump to create potential traps to prevent their decay~\cite{PhysRevLett.118.157401}. The principle is that the optical pumps in polariton systems not only provide the gain of the condensates, via exciton reservoirs, but also behave as potentials, strongly influencing the distribution of the condensates. For example, a 1D parabolic potential formed between two pump spots enables to observe phase-locked multiple condensates~\cite{PhysRevLett.110.186403,PhysRevLett.124.207402} and condensates occupying harmonic oscillator (SHO) states~\cite{tosi2012sculpting}. A 2D parabolic potential can lead to the formation of multilobe standing waves~\cite{PhysRevLett.107.106401} or vortices~\cite{lagoudakis2008quantized,ma2020realization}. In addition, the condensed polaritons simultaneously modulate the reservoir potentials, i.e., the optically induced potentials, which favors multistability of higher-order modes~\cite{ma2018vortex}. Higher-order dark solitons remain unexplored in nonlinear optics and atomic condensates, although multiple dark soliton interactions~\cite{PhysRevLett.96.043901,PhysRevLett.101.130401,becker2008oscillations} and dark-bright solitons~\cite{busch2001dark,becker2008oscillations,PhysRevLett.106.065302,PhysRevLett.128.033901} have been intensively studied. Dark double-hump solitons have been numerically demonstrated to be stable in higher-order nonlinear Schr{\"o}dinger equations~\cite{li2015dynamical,zhou2022generation}.

In this Letter, we use periodically modulated pumps to excite polariton condensates and find stable higher-order dark solitons trapped in the pump valleys. As the pump intensity increases gradually, we observe the relaxation of the dark solitons from the higher-order to the fundamental mode. During the relaxation, when several orders of dark states coexist in virtue of the  nonlinearity, their beating results in different oscillatory dynamics. If in each pump valley a dark soliton is excited, as the dark soliton squeezes the vicinal condensate, a distinct density minimum of the condensate can form between two dark solitons, i.e., in the pump peak area. We find that this density minimum is able to trap another type of dark state with a broader width and a fast oscillating behavior. In principle, the broader dark soliton is composed of two gray solitons.

\begin{figure}[t]
\centering
\fbox{\includegraphics[width=\linewidth]{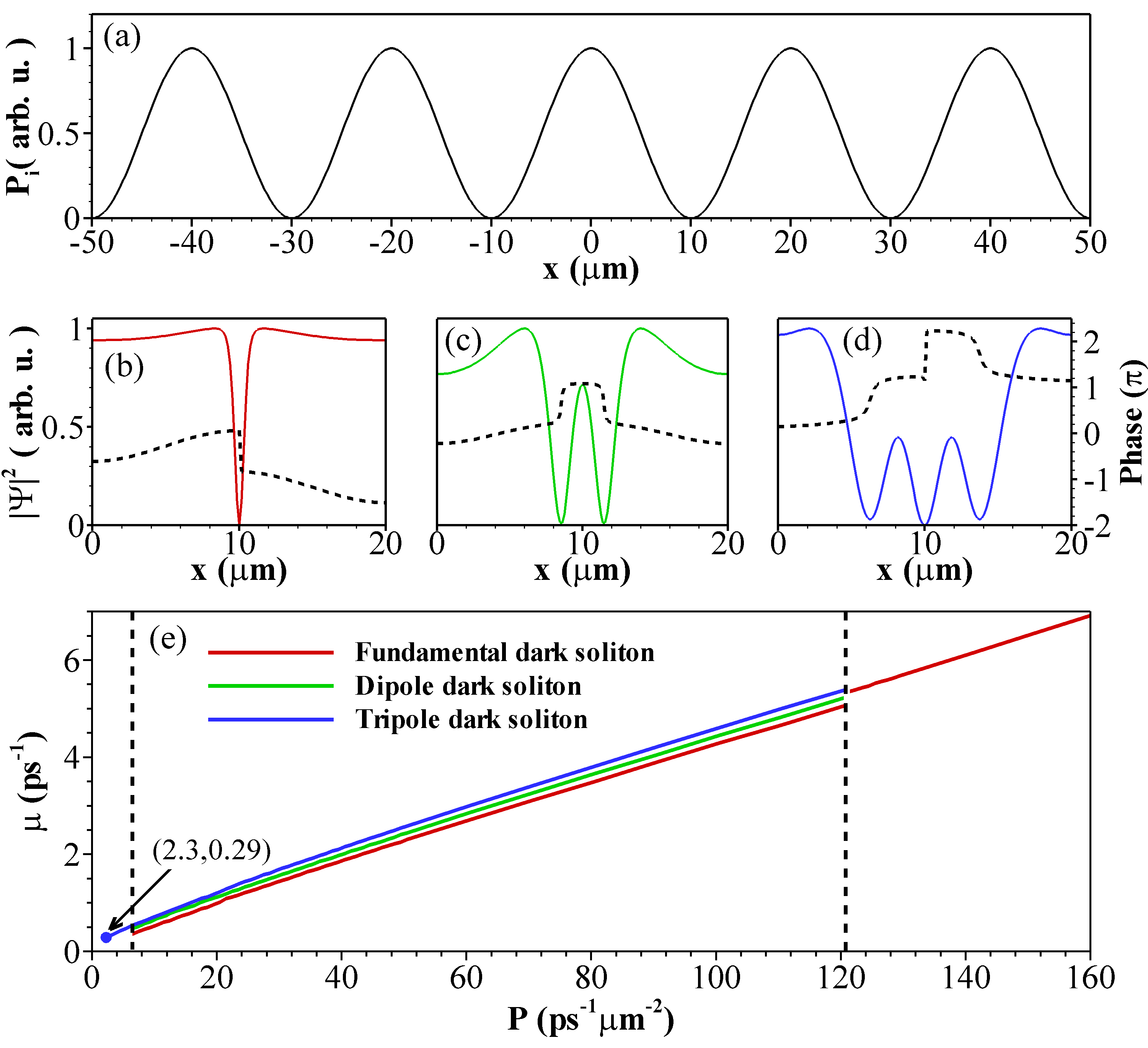}}
\caption{\textbf{Existence region of dark solitons.} (a) Distribution (normalized) of a periodic pump with the periodicity of $d=20$~$\mu$m. (b-c) Density (solid, normalized) and phase (dashed) profiles of the truncated (b) fundamental, (c) dipole, and (d) tripole dark solitons within one pump period at $P=130$~ps$^{-1}$~$\mu$m$^{-2}$, $P=20$~ps$^{-1}$~$\mu$m$^{-2}$, and $P=5$~ps$^{-1}$~$\mu$m$^{-2}$, respectively. (e) Dependence of the frequencies $\mu$ of the dark solitons on the pump intensity. The effective mass scaling parameter is $a=1$.}
\label{fig:1}
\end{figure}

The dynamics of a nonresonantly pumped polariton condensate in 1D can be described by a driven-dissipative Gross-Pitaevskii equation for the polariton field $\Psi(x,t)$, coupled with a rate equation describing the exciton reservoir $n(x,t)$ \cite{PhysRevLett.99.140402}, i.e.,
\begin{equation}\label{e1}
\begin{aligned}
i\hbar\frac{\partial\Psi(x,t)}{\partial t}&=\left[-\frac{\hbar^2}{2m}\frac{\partial^2 }{\partial x^2}-i\hbar\frac{\gamma_\textup{c}}{2}+g_\textup{c}|\Psi(x,t)|^2 \right.\\
&+\left.\left(g_\textup{r}+i\hbar\frac{R}{2}\right)n(x,t)\right]\Psi(x,t)\,,
\end{aligned}
\end{equation}
\begin{equation}\label{e2}
\frac{\partial n(x,t)}{\partial t}=\left[-\gamma_\textup{r}-R|\Psi(x,t)|^2\right]n(x,t)+P_\textup{i}(x,t)\,.
\end{equation}
Here, $m=10^{-4}m_\textup{e}/a$ is the effective mass of the polariton condensate which can be tuned by the constant $a$ and $m_\textup{e}$ is the free electron mass. \( \gamma_\textup{c}=0.1\,\mathrm{ps^{-1}} \mu m^2 \) and \( \gamma_\textup{r} = 1.5 \gamma_\textup{c} \) are the loss rates of the condensate and reservoir, respectively. $g_\textup{c} = 6 \times 10^{-3}\mathrm{meV} \mu m^2$ is the nonlinearity strength of the condensate and \( g_\textup{r} = 2g_\textup{c} \) represents the interaction between condensate and reservoir. $R = 0.01\,\mathrm{ps^{-1}} \mu m^2$ is the condensation rate. $P_\textup{i}(x)=
P\cos^2(\pi x/d)$ is a periodic incoherent pump with a period of $d= 20 \mu m$ and variable intensity $P$, see Fig.~\ref{fig:1}(a). To numerically solve the coupled Eqs.~(1) and (2), the Runge-Kutta (RK4) method has been applied with periodic boundary conditions, weak white initial noise for $\Psi(x,0)$, and zero initial conditions for $n(x,0)$.

\begin{figure}[t]
\centering
\fbox{\includegraphics[width=\linewidth]{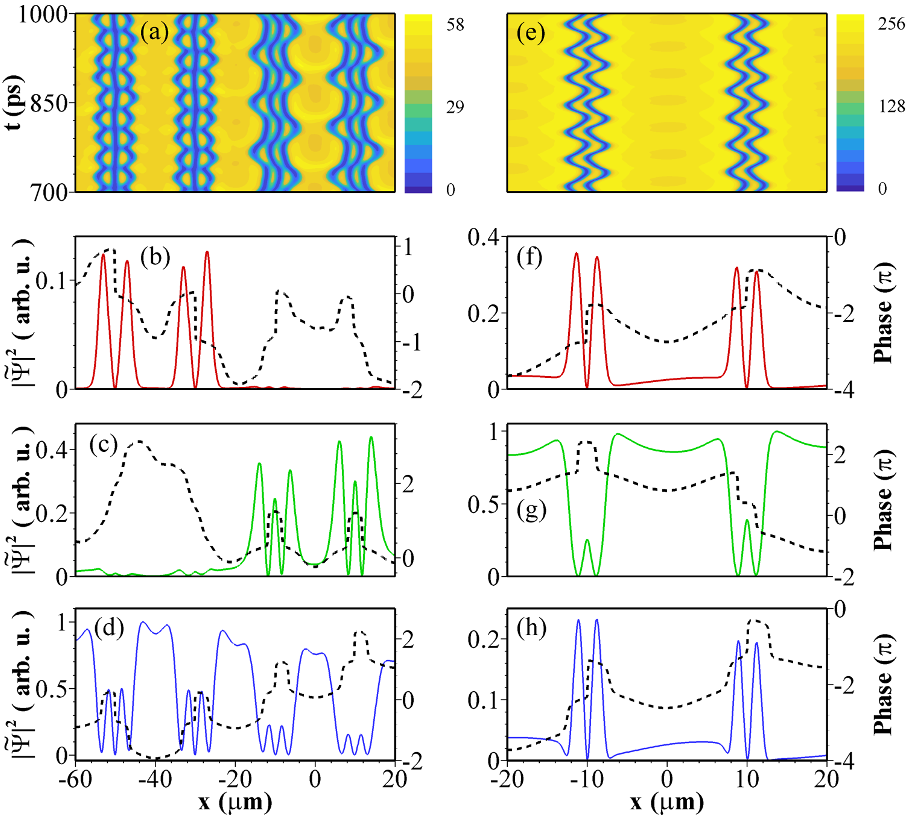}}
\caption{\textbf{Oscillatory dynamics.} (a) Time evolution of $|\Psi(x)|^2$ showing symmetric (left two) and asymmetric (right two) oscillators at $P=10$~ps$^{-1}$~$\mu$m$^{-2}$. Corresponding real-space spectral density (solid, normalized) and phase (dashed) profiles of (b) fundamental ($\mu=0.532$~ps$^{-1}$), (c) dipole ($\mu=0.637$~ps$^{-1}$), and (d) tripole ($\mu=0.72$~ps$^{-1}$) dark solitons. (e) Time evolution of $|\Psi(x)|^2$ showing two oscillators at $P=50$~ps$^{-1}$~$\mu$m$^{-2}$. Corresponding real-space spectral density (solid, normalized) and phase (dashed) profiles of (f) fundamental ($\mu=2.276$~ps$^{-1}$), (g) dipole ($\mu=2.414$~ps$^{-1}$), and (h) tripole ($\mu=2.552$~ps$^{-1}$) dark solitons. The effective mass scaling parameter is $a=1$.}
\label{fig:2}
\end{figure}

Under the excitation of the above-mentioned periodic pumps, different stationary solutions can occur at specific pump intensities. Figure \ref{fig:1} (b-d) presents three truncated stationary solutions (steady states) within one period (0-20 $\mu$m) of the pumps at $P=130$~ps$^{-1}$~$\mu$m$^{-2}$, $P=20$~ps$^{-1}$~$\mu$m$^{-2}$, and $P=5$~ps$^{-1}$~$\mu$m$^{-2}$, respectively. One can see that in the center of the pumps at $x=10$~$\mu$m, there are distinct density profiles and $\pi$-phase jumps, representing different orders of dark solitons, i.e., the fundamental dark soliton (FDS) [Fig. \ref{fig:1}(b)] with one pair of $\pi$-phase jump and density minimum, the dipole dark soliton (DDS) [Fig. \ref{fig:1}(c)] with two pairs of $\pi$-phase jumps and density minima, and the tripole dark soliton (TDS) [Fig. \ref{fig:1}(d)] with three pairs of $\pi$-phase jumps and density minima. All of them can be solely excited and stabilized in the same pump configuration but at different pump intensities. For example, the TDS is the only surviving state when the pump intensity is just above the condensation threshold ($P_\textup{th}=2.3{\leq}P{\leq}7$~ps$^{-1}$~$\mu$m$^{-2}$) as can be seen in Fig. \ref{fig:1}(e). For a stronger pumping intensity ($P>120$~ps$^{-1}$~$\mu$m$^{-2}$), however, the higher-order dark solitons vanish due to the deepening and narrowing of the pump-induced potential trap, leaving only the FDS stable [Fig. \ref{fig:1}(e)]. The stability of the steady states is demonstrated numerically by adding white noise at each picosecond during the time evolution up to 10 ns.

\begin{figure*}[h]
\centering
\fbox{\includegraphics[width=1\linewidth]{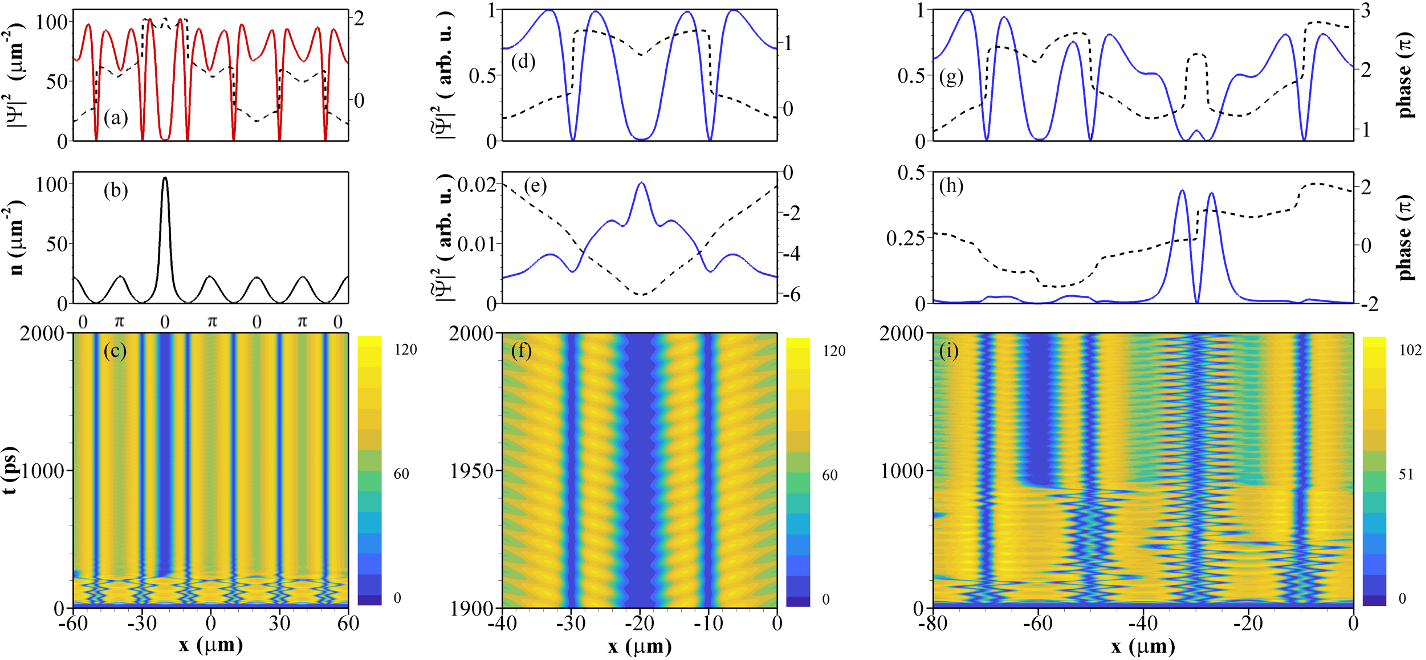}}
\caption{\textbf{Broader dark solitons at pump valleys.} (a) Density (solid) and phase (dashed) profiles of dark solitons and (b) their corresponding reservoir profiles collected after (c) the time evolution of $|\Psi(x)|^2$ at $t=2000$ ps, here $P=18$~ps$^{-1}$~$\mu$m$^{-2}$. The phase information above (c) indicate the phase jumps of the periodic background solution without the broader dark soliton. Real-space spectral density (solid, normalized) and phase (dashed) profiles with (d) $\mu=0.995$~ps$^{-1}$ and (e) $\mu=2.075$~ps$^{-1}$ of (f) the enlarged view of the selected time evolution in (c). Real-space spectral density (solid, normalized) and phase (dashed) profiles with (g) $\mu=0.876$~ps$^{-1}$ and (h) $\mu=0.750$~ps$^{-1}$ of the oscillatory dynamics in (i) the time evolution of $|\Psi(x)|^2$ at $P=15$~ps$^{-1}$~$\mu$m$^{-2}$. The effective mass scaling parameter is $a=2$.}
\label{fig:4}
\end{figure*}

For the pump intensity in between ($7<P{\leq}120$~ps$^{-1}$~$\mu$m$^{-2}$), all the three states can be simultaneously excited but with different amplitudes. Which state dominates during the excitation depends on the pump intensity as illustrated in Fig. \ref{fig:2}. When the pump intensity is close to the threshold, the TDS remains the strongest signal among them and exists in each pump valley [Fig. \ref{fig:2}(d)]. The FDS [Fig. \ref{fig:2}(b)] and DDS [Fig. \ref{fig:2}(c)] can also be triggered with much weaker densities, and their interactions with the TDS lead to different oscillatory dynamics, see Fig. \ref{fig:2}(a). As the pump intensity increases, the main signal transfers to the DDS, see Fig. \ref{fig:2}(f-h), and it oscillates under the influence of the signals from the FDS and DDS as time evolves, akin to the oscillatory dynamics of bright modes~\cite{PhysRevB.91.214301} but with inverted amplitudes. The spectral profiles [Figs. \ref{fig:2}(b-d) and \ref{fig:2}(f-h)] are obtained by (Fourier) transforming the oscillations from the time-domain to the frequency-domain, i.e., $\Psi(x,t){\longrightarrow}\widetilde{\Psi}(x,\mu)$. Further increasing the pump intensity results in further relaxation of the main signal to the ground state, until the higher-order states no longer exist. Note that the states at the pump intensity around the two dashed lines in Fig. \ref{fig:1}(e) appear or disappear gradually instead of abruptly. The density distributions in Fig. \ref{fig:2}(a,e) show that in this parameter combination the background densities are almost homogeneous without apparent modulations.

By slightly tuning the effective mass of the condensate, another kind of dark soliton is captured as shown in Fig. \ref{fig:4}(c) at $x=-20$ $\mu$m. Interestingly, this dark soliton resides in a pump peak area rather than a pump valley like the other surrounding narrower dark solitons. The appearance of this dark soliton is ascribed to the prominent density reduction of the condensate formed between two narrower dark solitons, see Fig. \ref{fig:4}(a). When the broader dark gap emerges, the impact of the condensate on the pump-induced reservoir via $R|\Psi|^2$ in \eqref{e2} vanishes. As a result, a stronger reservoir density at the same position is seen in Fig.~\ref{fig:4}(b) at $x=-20$ $\mu$m. On the other hand, the stronger reservoir density also induces a larger gain to the polariton condensate at that site which tends to kill the dark soliton. From the extracted states in the Fourier space one can see that apart from the dark soliton [Fig. \ref{fig:4}(d)], a density peak, akin to a bright soliton, can be clearly recognized in Fig. \ref{fig:4}(e), although its density is much weaker. The coexisting weaker bright peak perturbs the broader dark soliton to breath quickly in time with a symmetric distribution and a period of around 5.8 ps, see the enlarged time eovlution in Fig. \ref{fig:4}(f), which is much (approximately one order of magnitude) smaller than those in Fig. \ref{fig:2}. The amplitude of the bright signal in Fig. \ref{fig:4}(e) can be determined by the density of the reservoir, i.e., the pump intensity, and the width of the corresponding dark gap in Fig. \ref{fig:4}(d). However, a stronger bright signal may terminate the dark soliton. From the phase distribution of the bright signal in Fig. \ref{fig:4}(e), one can see that it propagates towards both directions, which can be identified from the inclined density stripes in Fig. \ref{fig:4}(f), and the outflowing dynamics hinder the formation of such broader dark solitons in the neighbouring pump peaks.

It is worth noting that the minimum density of this broader dark soliton approaches but is not exactly zero, since the phase difference besides the sharp density change is apparently smaller than $\pi$ as shown in Fig. \ref{fig:4}(d), which suggests that it is, strictly speaking, a gray soliton~\cite{pitaevskii2003bose}. It is known that different from dark solitons, a gray soliton carries a finite momentum so that it is not spatially pinned during the propagation or time evolution. The phase profile in Fig. \ref{fig:4}(d) also indicates that this broader dark gap is composed of two oppositely propagating gray solitons to maintain its localization. The simultaneous appearance of two gray solitons is determined by the phase-locked background condensate, see the phase information marked in Fig. \ref{fig:4}(c). These phase information represent the phase of the periodic solution that contains only the narrower dark solitons in each pump valley without the existence of the broader one (cf. the solution at $x>0$ in Fig.~\ref{fig:4}(a)). Therefore, to maintain the background $\pi$-phase jumps to avoid the abrupt change of the strong background density, which may require a strong energy, the appearance of the dark gap prefers to keep the symmetry of the phase on its both sides, allowing the emergence of twice phase jumps. In other words, the appearance of the broader dark soliton is unnecessary to strongly influence the background environment, with only the single cell it resides in being significantly reshaped, guaranteeing easy local manipulation of the broader dark soliton.

The bright signal occurring together with the broader dark soliton weakens until it disappears when the pump intensity is reduced. An example is shown in Fig. \ref{fig:4}(i) where the pump intensity is slightly reduced and a broader dark soliton is created at around 1000 ps and centered at $x=-60$ $\mu$m. The profile in Fig. \ref{fig:4}(g) shows that it consists of two gray solitons, while the corresponding signal at the lower state in Fig. \ref{fig:4}(h) is a weaker FDS instead of a bright one. This weaker FDS signal originates from the neighbouring irregular oscillation that is directly pumped from noise and it simultaneously triggers the broad dark soliton at around 1000 ps in Fig. \ref{fig:4}(i). With the participation of the antisymmetric phase of the weaker FDS, the solution oscillates around its center. As the pump intensity is reduced, higher-order dark solitons can also be excited in the pump valleys, such that another strong oscillation happens at $x=-30$ $\mu$m which contains both the DDS and FDS with nearly the same density, see Fig. \ref{fig:4}(g,h).

To conclude, we have studied two types of dark solitons in polariton condensates, supported by spatially periodically modulated pumping. One of them is narrower and trapped in the pump valleys in which the higher-order states can also be stabilized. In this case, we also observe different oscillatory dynamics when more than one dark state is selected. This sparks an interest in investigating dark simple harmonic oscillators. The other type of dark soliton has a broader size and can reside in the pump peaks when prominent density dips of the condensate in those areas are created. However, the broader dark soliton is not completely "dark", but composed of two opposite "gray" ones whose phase jumps are smaller than $\pi$. To stabilize the broader dark solitons, phase barriers imprinted by the narrower dark solitons in the pump valleys are essential. This broader dark soliton can coexist with a weaker bright signal with outgoing propagation and their interaction leads to the breathing of the broader dark soliton and the trembling dynamics of the surrounding narrower dark solitons. Such dynamics would inspire exploration on propagating bright solitons~\cite{PhysRevLett.102.153904,sich2012observation} boosted by dark solitons.

\begin{backmatter}
\bmsection{Funding} Deutsche Forschungsgemeinschaft (231447078, 270619725); Guangdong Basic and Applied Basic Research Foundation (2023A1515012432, 2024A1515010710).


\end{backmatter}



\begin{thebibliography}{10}
\newcommand{\enquote}[1]{``#1''}

\bibitem{kivshar2003optical}
Y.~S. Kivshar and G.~P. Agrawal, \emph{Optical solitons: from fibers to
  photonic crystals} (Academic press, 2003).

\bibitem{trillo2013spatial}
S.~Trillo and W.~Torruellas, \emph{Spatial solitons}, vol.~82 (Springer, 2013).

\bibitem{PhysRevLett.83.5198}
S.~Burger, K.~Bongs, S.~Dettmer, \emph{et~al.}, {\protect\JournalTitle{Phys.
  Rev. Lett.}} \textbf{83}, 5198 (1999).

\bibitem{denschlag2000generating}
J.~Denschlag, J.~E. Simsarian, D.~L. Feder, \emph{et~al.},
  {\protect\JournalTitle{Science}} \textbf{287}, 97 (2000).

\bibitem{becker2008oscillations}
C.~Becker, S.~Stellmer, P.~Soltan-Panahi, \emph{et~al.},
  {\protect\JournalTitle{Nature Physics}} \textbf{4}, 496 (2008).

\bibitem{amo2011polariton}
A.~Amo, S.~Pigeon, D.~Sanvitto, \emph{et~al.}, {\protect\JournalTitle{Science}}
  \textbf{332}, 1167 (2011).

\bibitem{PhysRevLett.107.245301}
G.~Grosso, G.~Nardin, F.~Morier-Genoud, \emph{et~al.},
  {\protect\JournalTitle{Phys. Rev. Lett.}} \textbf{107}, 245301 (2011).

\bibitem{PhysRevB.83.193305}
H.~Flayac, D.~D. Solnyshkov, and G.~Malpuech, {\protect\JournalTitle{Phys. Rev.
  B}} \textbf{83}, 193305 (2011).

\bibitem{PhysRevB.86.020509}
G.~Grosso, G.~Nardin, F.~Morier-Genoud, \emph{et~al.},
  {\protect\JournalTitle{Phys. Rev. B}} \textbf{86}, 020509 (2012).

\bibitem{PhysRevB.89.235310}
L.~A. Smirnov, D.~A. Smirnova, E.~A. Ostrovskaya, and Y.~S. Kivshar,
  {\protect\JournalTitle{Phys. Rev. B}} \textbf{89}, 235310 (2014).

\bibitem{PhysRevLett.112.216401}
Y.~Xue and M.~Matuszewski, {\protect\JournalTitle{Phys. Rev. Lett.}}
  \textbf{112}, 216401 (2014).

\bibitem{PhysRevLett.112.140405}
F.~Pinsker and H.~Flayac, {\protect\JournalTitle{Phys. Rev. Lett.}}
  \textbf{112}, 140405 (2014).

\bibitem{PhysRevLett.113.103901}
P.~Cilibrizzi, H.~Ohadi, T.~Ostatnicky, \emph{et~al.},
  {\protect\JournalTitle{Phys. Rev. Lett.}} \textbf{113}, 103901 (2014).

\bibitem{dominici2015real}
L.~Dominici, M.~Petrov, M.~Matuszewski, \emph{et~al.},
  {\protect\JournalTitle{Nature Communications}} \textbf{6}, 8993 (2015).

\bibitem{PhysRevLett.117.217401}
V.~Goblot, H.~S. Nguyen, I.~Carusotto, \emph{et~al.},
  {\protect\JournalTitle{Phys. Rev. Lett.}} \textbf{117}, 217401 (2016).

\bibitem{PhysRevLett.118.157401}
X.~Ma, O.~A. Egorov, and S.~Schumacher, {\protect\JournalTitle{Phys. Rev.
  Lett.}} \textbf{118}, 157401 (2017).

\bibitem{xue2018creation}
Y.~Xue, Y.~Jiang, G.~Wang, \emph{et~al.}, {\protect\JournalTitle{Optics
  Express}} \textbf{26}, 6267 (2018).

\bibitem{PhysRevX.10.041028}
A.~Ma\^{\i}tre, G.~Lerario, A.~Medeiros, \emph{et~al.},
  {\protect\JournalTitle{Phys. Rev. X}} \textbf{10}, 041028 (2020).

\bibitem{PhysRevB.105.245302}
J.~Wingenbach, M.~Pukrop, S.~Schumacher, and X.~Ma,
  {\protect\JournalTitle{Phys. Rev. B}} \textbf{105}, 245302 (2022).

\bibitem{PhysRevResearch.5.043062}
F.~Vercesi, Q.~Fontaine, S.~Ravets, \emph{et~al.}, {\protect\JournalTitle{Phys.
  Rev. Res.}} \textbf{5}, 043062 (2023).

\bibitem{PhysRevLett.110.186403}
P.~Cristofolini, A.~Dreismann, G.~Christmann, \emph{et~al.},
  {\protect\JournalTitle{Phys. Rev. Lett.}} \textbf{110}, 186403 (2013).

\bibitem{PhysRevLett.124.207402}
S.~Alyatkin, J.~D. T\"opfer, A.~Askitopoulos, \emph{et~al.},
  {\protect\JournalTitle{Phys. Rev. Lett.}} \textbf{124}, 207402 (2020).

\bibitem{tosi2012sculpting}
G.~Tosi, G.~Christmann, N.~Berloff, \emph{et~al.},
  {\protect\JournalTitle{Nature physics}} \textbf{8}, 190 (2012).

\bibitem{PhysRevLett.107.106401}
F.~Manni, K.~G. Lagoudakis, T.~C.~H. Liew, \emph{et~al.},
  {\protect\JournalTitle{Phys. Rev. Lett.}} \textbf{107}, 106401 (2011).

\bibitem{lagoudakis2008quantized}
K.~G. Lagoudakis, M.~Wouters, M.~Richard, \emph{et~al.},
  {\protect\JournalTitle{Nature physics}} \textbf{4}, 706 (2008).

\bibitem{ma2020realization}
X.~Ma, B.~Berger, M.~A{\ss}mann, \emph{et~al.}, {\protect\JournalTitle{Nature
  Communications}} \textbf{11}, 897 (2020).

\bibitem{ma2018vortex}
X.~Ma and S.~Schumacher, {\protect\JournalTitle{Physical Review Letters}}
  \textbf{121}, 227404 (2018).

\bibitem{PhysRevLett.96.043901}
A.~Dreischuh, D.~N. Neshev, D.~E. Petersen, \emph{et~al.},
  {\protect\JournalTitle{Phys. Rev. Lett.}} \textbf{96}, 043901 (2006).

\bibitem{PhysRevLett.101.130401}
A.~Weller, J.~P. Ronzheimer, C.~Gross, \emph{et~al.},
  {\protect\JournalTitle{Phys. Rev. Lett.}} \textbf{101}, 130401 (2008).

\bibitem{busch2001dark}
T.~Busch and J.~Anglin, {\protect\JournalTitle{Physical review letters}}
  \textbf{87}, 010401 (2001).

\bibitem{PhysRevLett.106.065302}
C.~Hamner, J.~J. Chang, P.~Engels, and M.~A. Hoefer,
  {\protect\JournalTitle{Phys. Rev. Lett.}} \textbf{106}, 065302 (2011).

\bibitem{PhysRevLett.128.033901}
S.~Zhang, T.~Bi, G.~N. Ghalanos, \emph{et~al.}, {\protect\JournalTitle{Phys.
  Rev. Lett.}} \textbf{128}, 033901 (2022).

\bibitem{li2015dynamical}
M.~Li, T.~Xu, and L.~Wang, {\protect\JournalTitle{Nonlinear Dynamics}}
  \textbf{80}, 1451 (2015).

\bibitem{zhou2022generation}
Q.~Zhou, M.~Xu, Y.~Sun, \emph{et~al.}, {\protect\JournalTitle{Nonlinear
  Dynamics}} \textbf{110}, 1747 (2022).

\bibitem{PhysRevLett.99.140402}
M.~Wouters and I.~Carusotto, {\protect\JournalTitle{Phys. Rev. Lett.}}
  \textbf{99}, 140402 (2007).

\bibitem{PhysRevB.91.214301}
X.~Ma, I.~Y. Chestnov, M.~V. Charukhchyan, \emph{et~al.},
  {\protect\JournalTitle{Phys. Rev. B}} \textbf{91}, 214301 (2015).

\bibitem{pitaevskii2003bose}
L.~Pitaevskii and S.~Stringari, \enquote{Bose-einstein condensation,
  clarendon,}  (2003).

\bibitem{PhysRevLett.102.153904}
O.~A. Egorov, D.~V. Skryabin, A.~V. Yulin, and F.~Lederer,
  {\protect\JournalTitle{Phys. Rev. Lett.}} \textbf{102}, 153904 (2009).

\bibitem{sich2012observation}
M.~Sich, D.~Krizhanovskii, M.~Skolnick, \emph{et~al.},
  {\protect\JournalTitle{Nature Photonics}} \textbf{6}, 50 (2012).

\end{thebibliography}


\end{document}